\documentstyle[12pt]{article}
\newcommand{\bold}[1]{\mbox{\boldmath ${#1}$}}
\begin{document}
\input{epsf}
\input{psfig}
\baselineskip 4 ex
\title{Delta-Hole Interaction in Nuclei and the Gamow-Teller Strength
in $^{90}$Nb 
\thanks{E-mail addresses: bentz@keyaki.cc.u-tokai.ac.jp (W. Bentz),
suzuki@chs.nihon-u.ac.jp (Toshio Suzuki), suzuki@quantum.apphy.fukui-u.ac.jp
(Toshio Suzuki)}}  
\author{Akito Arima \\
House of Councillors \\
2-1-1 Nagata-cho, Tokyo 100-8962, Japan \\
{ } \\
Wolfgang Bentz \\
Dept. of Physics, School of Science, \\
Tokai University \\
1117 Kita-Kaname, Hiratsuka 259-1207, Japan \\
{ }\\
Toshio Suzuki \\
Dept. of Physics, College of Humanities and Sciences, \\
Nihon University \\
3-25-40 Sakurajosui, Setagaya-ku, Tokyo 156, Japan \\
{ }\\
and \\
{ }\\
Toshio Suzuki \\
Dept. of Applied Physics, Faculty of Engineering, \\
Fukui University \\
3-9-1 Bunkyo, Fukui 910-8507, Japan \\
and\\
RIKEN\\
2-1 Hirosawa, Wako-shi, Saitama 351-0198, Japan}

\date{}
\maketitle
\newpage
\begin{abstract}

The recently obtained experimental result for the quenching 
of the GT sum rule in the reaction $^{90}$Zr(p,n)$^{90}$Nb is used to extract the value of the
Landau-Migdal parameter $g_{{\rm N}\Delta}'$, taking into account also the effect
of the finite range meson exchange interactions. The
extracted value is compared to the  
one obtained in the $\pi+\rho$ exchange model by explicitly taking into account the
effects of antisymmetrization and short range correlations. Although the $\pi$+$\rho$ exchange model 
tends to give a somewhat stronger quenching than observed experimentally, the results are
consistent within the experimental error bars if the quark model value for the parameter
$f_{\Delta}/f_{\pi}$ is used.\\ 

Keywords; Landau-Migdal parameters; Quenching of Gamow-Teller strength;
          $\Delta$-hole interaction 

\end{abstract}

The study of the renormalization of Gamow-Teller (GT) matrix elements in nuclei
has revealed valuable information on nuclear structure effects
and non-nucleonic degrees of freedom like mesons and delta-isobars \cite{ASB,MIG}. If one
considers only the low energy part of the GT strength function, as provided by the
GT $\beta$-decay matrix elements or the charge exchange reactions at low excitation 
energies, one cannot decide {\em a priori} whether the
experimentally observed quenchings originate from nuclear structure effects, like
configuration mixing and core polarization, or from non-nucleonic
degrees of freedom. Although theoretical calculations with realistic interactions
have shown that a considerable amount of strength is shifted to the higher energy region due to the 
admixture of $2p-2h$ components in even-even nuclei\cite{BER}, the observed quenchings have also been attributed
to the admixture of $\Delta$-hole components by assuming a large value of the
Landau-Migdal parameter $g_{{\rm N}\Delta}'$ characterizing the strength of the $NN\rightarrow
N\Delta$ transition potential \cite{DELT}. In order to discriminate between these two interpretations,
Wakasa et al. \cite{WAK} have recently used the reaction $^{90}$Zr(p,n)$^{90}$Nb to extract the value
of the GT sum rule (or more precisely the Ikeda sum rule \cite{IKE}) integrated up to about $50$ MeV 
excitation energy. Since this value
of the excitation energy is sufficiently high to include most, if not all, of the 
strength due to $2p-2h$ excitations \cite{DAN}, their result $(S_- -S_+)_{\rm exp}/S_{-0}=0.9 \pm 0.05$ indicates
that the quenching due to the $\Delta$-hole excitations should be about $10\%$ or even smaller.
\footnote{Here $S_{\mp}$ is the sum rule for the GT transition $(N,Z)\rightarrow (N \mp 1,Z \pm 1)$. In the 
convention used here the neutron (proton) has $t_z=1/2 \,\,(t_z=-1/2)$. The Ikeda sum rule 
$S_- -S_+=3(N-Z)$ is satisfied by the single particle values, i.e; $S_{-0}=3(N-Z),\,\,S_{+0}=0$.}

In order to extract information on the value of $g_{{\rm N}\Delta}'$ from this observation,
Suzuki and Sakai \cite{TS} recently performed an analysis by assuming that the relevant part of the 
transition potential is solely given by the delta-function type Landau-Migdal interaction.
For the case of vanishing $N\Delta\rightarrow N\Delta$ interaction, their analysis gave  
$g_{{\rm N}\Delta}'=0.18$ ($g_{{\rm N}\Delta}'=0.12$) if the quark model (Chew-Low model) value of
the parameter $f_{\Delta}/f_{\pi}$ (see eqs. (1) and (2) below) is assumed, which is considerably smaller than one
would expect on the basis of the $\pi+\rho$ exchange model \cite{ACS}.  
The assumption of a pure delta function potential is justified in nuclear matter, 
since the direct term due to the 
'bare' $\pi$ + $\rho$ meson exchange transition potential ($V_{\pi+\rho}(\bold{q}$))
vanishes for ${\bold q}=0$. In finite nuclei, however, due to the lack of momentum
conservation, $V_{\pi+\rho}(\bold{q})$ acts attractively and enhances the GT matrix elements,
which would imply a larger value of $g_{{\rm N}\Delta}'$ in order to balance this attraction. 
The purpose of this letter is to investigate this possibility and to comment on the
question whether the $\pi+\rho$ model is consistent with the
experimentally observed quenching or not.

Following the perturbative treatment of Arima et al \cite{ACS}, the GT sum rule $S_-$ including
the effects of $\Delta$-hole excitations in lowest order is represented diagrammatically
by fig.1. 

\begin{figure}[hbt]
\begin{center}
\psfig{file=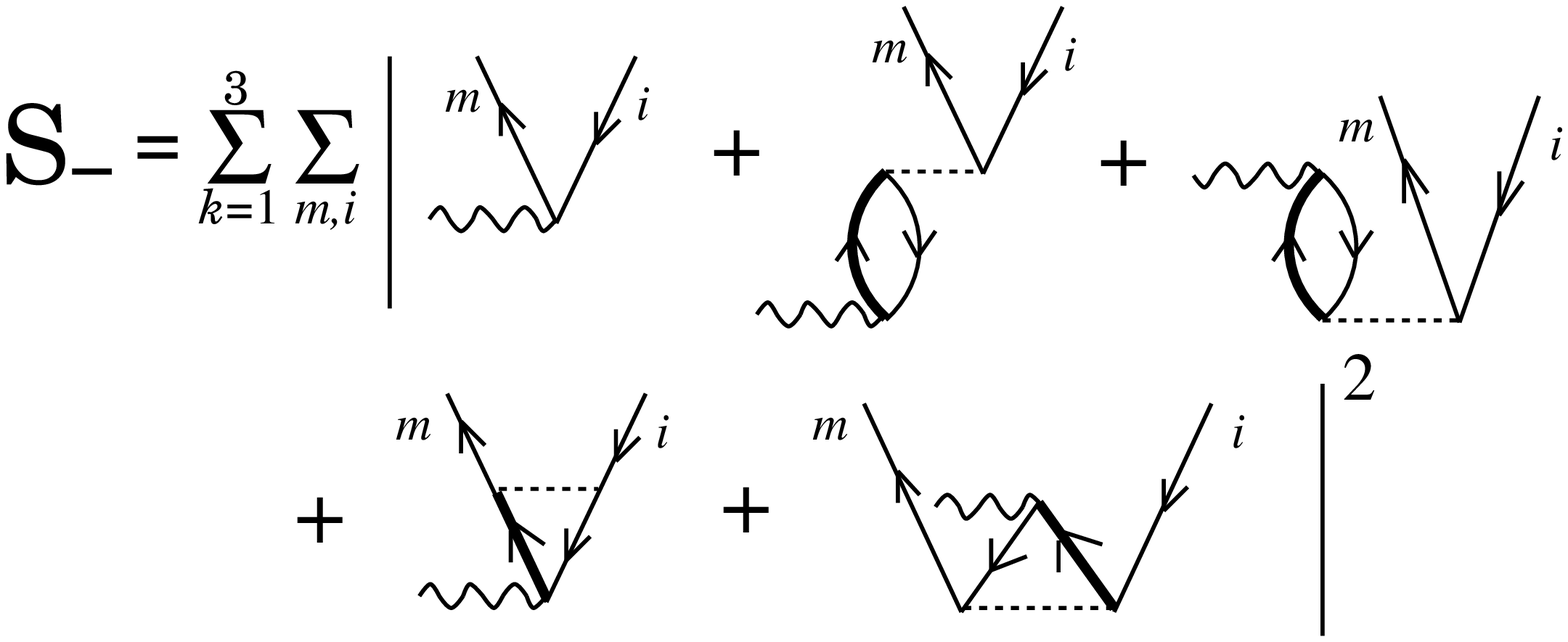,height=2.5in}
\caption{Diagrammatical representation of the GT sum rule including the effects of $\Delta$h-states
in lowest order perturbation theory.}
\end{center}
\end{figure}

The external GT operator is given in the space of $N$ and $\Delta$ states by
\begin{equation}
{\cal O}_{k-}=\frac{1}{2}\left(\sigma_k \,\,\tau_- + \frac{f_{\Delta}}{f_{\pi}} S_k \,\,T_-  
+ \frac{f_{\Delta}}{f_{\pi}} (S^{\dagger})_k\,\, (T^{\dagger})_- \right), 
\label{op}
\end{equation}
where the transition
spin ${\bold S}$ (transition isospin ${\bold T}$) transforms a nucleon into a delta 
with the reduced matrix elements given by $<\frac{3}{2}||S||\frac{1}{2}>=
<\frac{3}{2}||T||\frac{1}{2}>=2$ \footnote{We define the spherical $\pm$ components by 
$A_{\pm}=A_1 \pm i A_2$. Useful relations for the transition spin are $(S^{\dagger})_i S_j=
\frac{2}{3} \delta_{ij} -\frac{i}{3} \epsilon_{ijk} \sigma_k$ and similar for the transition 
isospin. } We will use the quark model value $\frac{f_{\Delta}}{f_{\pi}}=
\sqrt{\frac{72}{25}}$ as in ref. \cite{TS}, and also $\frac{f_{\Delta}}{f_{\pi}}=2$ \cite{ACS},
which is based on the Chew-Low model \cite{CR}. For the transition $^{90}$Zr$\rightarrow ^{90}$Nb, 
the relevant proton-particle
neutron-hole states $(m,i)$ in fig. 1 are $(g_{9/2},g_{9/2})$ and $(g_{7/2},g_{9/2})$.
The first diagram alone gives the single particle value $S_{-0}=3(N-Z)=30$.
The diagrams of fig.1 refer to a calculation in which the effects of the exchange terms
and short range correlations (s.r.c.) are explicitly taken into account, that is, in the $\pi$+
$\rho$ exchange model the transition potential is given by $V_{\pi+\rho}(\bold{r})f(r)$, where
$f(r)$ is a s.r.c. function \footnote{In the actual calculation we use 
$f(r)=\Theta(r-c)$ with $c=0.7$ fm.}, and the 'bare' transition potential is (in momentum space)
\begin{eqnarray}
V_{\pi+\rho}(\bold{q})&=& - \frac{f_{\Delta}}{f_{\pi}} \left(\left(\frac{f_{\pi}}{m_{\pi}}\right)^2
\frac{\bold{\sigma}_1 \cdot {\bold q}\,\, \bold{S}_2 \cdot \bold{q}}{\bold{q}^2+m_{\pi}^2} + 
\left(\frac{f_{\rho}}{m_{\rho}}\right)^2
\frac{\left(\bold{\sigma}_1 \times {\bold q}\right) \cdot \left(\bold{S}_2 \times \bold{q}\right)}
{\bold{q}^2+m_{\rho}^2} \right) \bold{\tau}_1 \cdot \bold{T}_2 \nonumber \\
&+& (1 \leftrightarrow 2) + h.c.
\label{pot}
\end{eqnarray}
Here we use the coupling constants $f_{\pi}^2=4\pi \cdot 0.08 = 1.01$, and $f_{\rho}^2 = 
4\pi \cdot 4.86 = 61.04$ \cite{HP}. We use harmonic oscillator single particle wave functions
with $\hbar \omega=9.23$ MeV (b=$2.12$ fm) \cite{VJV}.
Note that in a calculation of this type, which includes explicitly the exchange terms 
and s.r.c., $g_{{\rm N}\Delta}'$ does not appear explicitly.
We will refer to such a kind of calculation as 'type 1'. 

The resulting quenching factors $S_-/S_{-0}$ obtained in this type 1 calculation
are shown in the second and fourth columns of Table 1.
Here '$\pi$ only' refers to a calculation taking only the pion exchange potential
($V_{\pi}$) supplemented by the s.r.c. function, and '$\pi+\rho$'
includes also the effects of $\rho$ meson exchange.
Comparing these results to the experimental value of ref. \cite{WAK}, we see that, for the 
case of the quark model value $\,\,\frac{f_{\Delta}}{f_{\pi}}=\sqrt{\frac{72}{25}}$, pion exchange alone
leads to the observed quenching factor, but the quenching obtained by including
also the $\rho$ meson exchange is somewhat too strong, although it is still within the
error bars of the experimental value. The quenching obtained with $\,\,\frac{f_{\Delta}}{f_{\pi}}=2$,
however, seems to be too strong compared to the experimental value. 

\begin{table}[h]
\begin{center}
\begin{tabular}{|l|l|l||l|l|} 
\hline
 & \multicolumn{2}{c||}{(a)$\,\,\frac{f_{\Delta}}{f_{\pi}}=\sqrt{\frac{72}{25}}$} 
 &\multicolumn{2}{c|}{(b)$\,\,\frac{f_{\Delta}}{f_{\pi}}=2$} \\  
 \cline{2-5} & $S_+/S_{0+}$ & $g_{{\rm N}\Delta}'$   & $S_+/S_{0+}$ & $g_{{\rm N}\Delta}'$ \\  
 \cline{2-5} $\pi$ only &  0.90 & 0.25  & 0.867 & 0.25 \\
 $\pi+\rho$  & 0.86 & 0.35 & 0.805 & 0.35\\  \hline 
\end{tabular}
\end{center}
\caption{Results for the quenching factors and corresponding values of $g_{{\rm N}\Delta}'$, referring
to the two cases (a)$\,\,\frac{f_{\Delta}}{f_{\pi}}=\sqrt{\frac{72}{25}}$, and
(b)$\,\,\frac{f_{\Delta}}{f_{\pi}}=2$.   
The quenching factors shown here are obtained in a calculation including
exchange terms and short range correlations explicitly ('type 1'). The corresponding
values of $g_{{\rm N}\Delta}'$ are determined such that a calculation ('type 2') without the exchange terms and 
short range correlation function, but employing the $g_{{\rm N}\Delta}'$ force in addition to the
bare meson exchange potentials ($V_{\pi}$ and $V_{\pi+\rho}$), gives the same quenching factors
as the type 1 calculation. 
The observed quenching factor is ref. \cite{WAK} $0.9 \pm 0.05$. }
\end{table}

In order to translate these results into values for $g_{{\rm N}\Delta}'$, we also performed a calculation
of 'type 2', which considers only the direct terms in fig.1  with the bare meson exchange 
potentials (without s.r.c. functions), but supplemented by the contact interaction of Landau-Migdal
type (in momentum space),

\begin{equation}
V_{L.M.}= g_{{\rm N}\Delta}' \left(\frac{f_{\pi}}{m_{\pi}}\right)^2 \frac{f_{\Delta}}{f_{\pi}}
\left(\bold{\sigma}_1 \cdot \bold{S}_2 \,\,\bold{\tau}_1 \cdot \bold{T}_2 + (1 \leftrightarrow 2) + 
{\rm h.c.} \right)
\label{lm}      
\end{equation}

In this 'type 2' calculation, the effects of the exchange terms and s.r.c. are incorporated
into the interaction (\ref{lm}). If we determine the value of $g_{{\rm N}\Delta}'$ such as to reproduce
the results of the type 1 calculation, we obtain the values shown in table 1.
The resulting value $g_{{\rm N}\Delta}'=0.35$ shown in table 1 is very similar to the average value deduced in ref. \cite{ACS}
for finite nuclei. (The value obtained in nuclear matter for normal density is somewhat larger,
around $0.4$.) 

On the other hand, we can also determine 'empirical' values of $g_{{\rm N}\Delta}'$ such
as to reproduce the experimental quenching factor. The results are shown in Table 2. They are
obtained by assuming (i) only the Landau-Migdal type interaction (\ref{lm}),
(ii) the interaction (\ref{lm}) together with $V_{\pi}$, and (iii) the interaction (\ref{lm})
together with $V_{\pi+\rho}$. These results, of course, refer to the 'type 2' calculation, i.e; only the
direct terms of fig. 1 are considered and no s.r.c. function is used. 

\begin{table}[h]
\begin{center}
\begin{tabular}{|l|l||l|} 
\hline
& (a)$\,\, \frac{f_{\Delta}}{f_{\pi}}=\sqrt{\frac{72}{25}}$ & (b)$\,\,\frac{f_{\Delta}}{f_{\pi}}=2$ \\
\hline
$V_{L.M.}$ only & 0.18 $\pm$ 0.09  & 0.13 $\pm$ 0.07 \\
$V_{L.M.}+V_{\pi}$ & 0.26 $\pm$ 0.09 & 0.20 $\pm$ 0.07 \\
$V_{L.M.}+V_{\pi+\rho}$ & 0.27 $\pm$ 0.09 & 0.21 $\pm$ 0.07 \\
\hline
\end{tabular}
\end{center}
\caption{Values of $g_{{\rm N}\Delta}'$ obtained by fitting the experimental quenching factor 
assuming various types of interactions (see text).}
\end{table} 

The value $g_{{\rm N}\Delta}'=0.18$ ($g_{{\rm N}\Delta}'=0.13$) for the case of the quark model 
(Chew-Low model) value of $\frac{f_{\Delta}}{f_{\pi}}$ agrees with the result of ref. \cite{TS}, 
where only the interaction
(\ref{lm}) was used. (A small difference arises because $\frac{f_{\Delta}}{f_{\pi}}=2.12$ was
used for the Chew-Low model value in ref. \cite{TS}.) Since the 'bare' meson exchange potentials (\ref{pot}) vanish for $q=0$, 
their direct terms give no contribution in nuclear matter due to momentum conservation, and
the assumption of a pure $\delta$-function type interaction would be justified in nuclear matter. 
However, we see
from Table 2 that the attractive contribution due to pion exchange is appreciable in finite
nuclei, and in order to balance the attraction due to $V_{\pi}$ the value of $g_{{\rm N}\Delta}'$
has to be increased. The $\rho$ meson exchange contributes only little, since it is of short range
and therefore the nuclear matter picture is valid to a good approximation. From Table 2 we can
conclude that, due to the presence of $V_{\pi+\rho}$, the value of $g_{{\rm N}\Delta}'$ 
increases by about 0.1 and, for the case of the quark model $(\frac{f_{\Delta}}{f_{\pi}}=\sqrt{\frac{72}{25}}$),
becomes consistent with the meson exchange
picture (Table 1). For the case of the Chew-Low model $(\frac{f_{\Delta}}{f_{\pi}}=2$),
however, the value of $g_{{\rm N}\Delta}'$ fitted to the experimental quenching is definitely
smaller than the value obtained in the meson exchange picture. This corresponds to the
situation discussed above that the quenching calculated in the meson exchange picture
is consistent with (stronger than) the experimental quenching for the case of the quark model (Chew-Low model)
value of $\frac{f_{\Delta}}{f_{\pi}}$.  

It is interesting to note that our present perturbative approach is equivalent to the
RPA formulation of ref. \cite{TS} for the case of vanishing $N\Delta\rightarrow N\Delta$ interaction.
In order to
see the connection clearly, let us derive the analytical result for the quenching factor
obtained in ref. \cite{TS} from the diagrams of fig.1: As usual, one can approximate the energy denominators
by a constant ($-1/\epsilon_{\Delta}$ with $\epsilon_{\Delta}\simeq 294$ MeV), which allows one 
to perform the sum over the $\Delta$ states by using completeness. In this way one obtains
from fig.1
\begin{eqnarray}
S_- &=& \frac{1}{4} \sum_{k=1}^{k=3} \sum_{mi} |\langle m| \sigma_k \,\,\tau_-|i \rangle
- g_{{\rm N}\Delta}'\left(\frac{f_{\pi}}{m_{\pi}}\frac{f_{\Delta}}{f_{\pi}}\right)^2 \frac{8}
{9\epsilon_{\Delta}} \nonumber \\
&\times& \sum_h \langle h|\left(\delta_{kl} - \frac{i}{4} \,\,\tau_3 \,\,\epsilon_{kk'l} \,\,
\sigma_{k'}\right)\,\,G_{mi}^{(l)}({\bold r_h})|h \rangle|^2 
\label{sum}
\end{eqnarray}
with the spin-isospin density
\begin{equation}
G_{mi}^{(l)}({\bold r})=\langle m| \sigma_l \,\,\tau_- \delta({\bold r}-{\bold r_i})|i \rangle.
\end{equation}
By using simple angular momentum algebra one can show that the spin dependent term in eq.(\ref{sum})
is zero. Then the term $\langle h|...|h\rangle$ in (\ref{sum}) becomes 
$\int d^3 r \rho_0({\bold r}) \,\,G_{mi}^{(k)}({\bold r})$, where 
$\rho_0({\bold r})=\sum_h \langle h| \delta({\bold r_h}-{\bold r})|h\rangle$ is the ground state nucleon density
of $^{90}$Zr. Since $G_{mi}^{(k)}({\bold r})$ is strongly
peaked at the nuclear surface, one can approximate this integral by replacing 
$\rho_0({\bold r})$ by its value at the nuclear surface, i.e; $\rho_0({\bold r}) \rightarrow \gamma \rho_0$,
where $\gamma\simeq 0.5$ is the attenuation factor for nuclear surface effects, and $\rho_0=0.17$ fm$^{-3}$.
Since the remaining integral over the spin-isospin density just gives the single-particle matrix element
$\langle m| \sigma_k \,\,\tau_-|i \rangle$, one ends up with a state independent quenching factor
$Q\equiv S_-/S_{-0}$ given by
\begin{equation}
Q=\left(1-g_{{\rm N}\Delta}'\left(\frac{f_{\pi}}{m_{\pi}}\frac{f_{\Delta}}{f_{\pi}}\right)^2 \frac{8}
{9\epsilon_{\Delta}} \gamma \rho_0\right)^2,
\label{quen}
\end{equation}
and with the quark model value $\left(\frac{f_{\Delta}}{f_{\pi}}\right)^2=\frac{72}{25}$, this
corresponds to eq.(35) of ref. \cite{TS} for the case $g_{\Delta \Delta}'=0$. 
\footnote{We note that the term   
$(Z-N)/2A$ in eq. (36) of ref. \cite{TS} should actually be absent due to a cancellation between the forward
and backward $\Delta h$ contributions. However, this term is very small for the present case of
$^{90}$Zr. It is actually overwhelmed by ambiguities in the attenuation factor $\gamma$, such that
our numerical results for the case '$V_{L.M.}$ only' shown in Table 2 are actually the same as in ref. \cite{TS}.}
The values $g_{{\rm N}\Delta}'=0.18$ and $g_{{\rm N}\Delta}'=0.13$ listed in table 2, which have been obtained 
without using the approximation for nuclear surface effects, correspond to $\gamma=0.49$.

In conclusion, we have shown that the recently obtained experimental result for the quenching
of the GT sum rule in $^{90}$Nb is consistent with the meson exchange picture for the 
$\Delta h$ interaction in nuclei, if the quark model value for $\frac{f_{\Delta}}{f_{\pi}}$ is assumed. 
In order to arrive at this conclusion, it is important
to take into account also the finite range $\pi+\rho$ exchange potentials in addition to the Landau-Migdal
type interaction.  

\vspace{1 cm}

{\sc Acknowledgement}\\
The authors wish to thank Hideyuki Sakai for his encouragement and helpful discussions.

\end{document}